\documentclass[11pt,twoside]{article}
\usepackage{./asp2014}

\include{graphics}

\aspSuppressVolSlug
\resetcounters

\begin{document}

\title{White Dwarfs in the Metal-Rich Open Cluster NGC 6253}
\author{Elizabeth J. Jeffery,$^1$ Fabiola Campos,$^2$ Alejandra Romero,$^{3}$ and S.O. Kepler,$^4$}
\affil{$^1$Brigham Young University, Provo, UT, USA; \email{ejeffery@byu.edu}}
\affil{$^2$University of Texas at Austin, Austin, TX, USA; \email{fcampos@astro.as.utexas.edu}}
\affil{$^3$UFRGS, Porto Alegre, Brazil; \email{alejandra.romero@ufrgs.br}}
\affil{$^4$UFRGS, Porto Alegre, Brazil; \email{kepler@ufrgs.br}}

\paperauthor{Elizabeth J. Jeffery}{ejeffery@byu.edu}{}{Brigham Young University}{Department of Physics and Astronomy}{Provo}{UT}{84602}{USA}

\begin{abstract}
We have obtained 53 images with the $g$ filter and 19 images with the $i$ filter, each with 600-second exposures of the super metal rich open cluster NGC 6253 with the Gemini-South telescope to create deep images of the cluster to observe the cluster white dwarfs for the first time. We will analyze the white dwarf luminosity function to measure the cluster's white dwarf age, search for any anomalous features (as has been seen in the similarly metal rich cluster NGC 6791), and constrain the initial-final mass relation at high metallicities. We present an update on these observations and our program to study the formation of white dwarfs in super high metallicity environments.
\end{abstract}

\section{Scientific Motivation and Background}
\label{intro}
   Understanding the effect of metallicity on white dwarf (WD) cooling has applications to many aspects of stellar evolution, including the initial-final mass relation (IFMR) and using WDs to measure the ages of stellar populations. The unusual WD luminosity function (LF) of the metal-rich open cluster NGC 6791 (Bedin et al. 2008a) and subsequent attempts to explain it (e.g., Hansen 2005; Bedin et al. 2008b; Garcia-Berro et al. 2009) has motivated us to study the WDs in other metal-rich clusters.

   NGC 6253 has been found to be just as metal rich as NGC 6791 ([Fe/H] $\sim$ +0.43, Anthony-Twarog et al. 2010), but it is not as old. Two different ages for NGC 6253 have been reported in recent studies in the literature. Montalto et al. (2009) measured proper motions of stars in the field of NGC 6253 and fit Padova and Yale-Yonsei isochrones to the main sequence turn off (MSTO) on the color-magnitude diagram (CMD) of likely cluster members and estimated a cluster age of 3.5 Gyr. Rozyczka et al. (2014) determined the mass of an eclipsing binary star at the cluster TO, and from this measured a cluster age of 4.6 Gyr.
   
   We have obtained deep images of NGC 6253 to observe the cluster WDs for the first time. In these proceedings we will describe the observations, methods, and present preliminary results.

\section{Observations and Photometric Calibrations}
\label{obs}

   For this study, we have used two different data sets. In this section we describe these data and the data reduction process. Here we also discuss data reduction and methods for photometric calibrations.
   
\subsection{Cluster Images}
\label{images}

   We have used the imaging mode on the Gemini MultiObject Spectrograph (GMOS) on the Gemini-South telescope to obtain observations of NGC 6253. We have secured short, medium, and long exposures in the $g$ and $i$ filters. The long observations consist of 19 $i$ (3.2 hours) and 55 $g$ (9.1 hours) images; all $i$ images and 32 $g$ images are included in the analysis here.
   
   In addition to the GMOS images, we have taken images of both the cluster and standard fields (Smith et al. 2007) in $gri$ filters using the Southeastern Association for Research in Astronomy (SARA) 0.6-meter telescope at Cerro Tololo Interamerican Observatory (CTIO). These observations were used to transform stars in the cluster field to the standard system. By determining the standard magnitudes of this subset of stars in the cluster field, we used these stars as local standards on the GMOS images. This allows us to transform the Gemini on to the standard system.
   
\subsection{Data Reductions}
\label{data_red}

   The SARA images were reduced using standard procedures in IRAF. Instrumental magnitudes for both cluster images and standard fields were measured using the IRAF PHOT package. Instrumental magnitudes were transformed to the standard system using the tools within the IRAF PHOTCAL package. For stars with multiple observations, we averaged results to obtain a single, standardized magnitude.

   GMOS images were reduced using the software package THELI\footnote{url for theli here} (Schirmer 2013; Erben et al. 2005), including bias and flat fielding. The output of THELI is a science-ready, mosaic image. We processed the short, medium, and long images separately, creating three master images. To determine instrumental magnitudes, we used point spread function (PSF) methods, utilizing the DAOPHOT/ALLSTAR (Stetson 1987) software. 

   By matching stars that appeared on both the cluster images taken with SARA and GMOS, we were able to determine the photometric offset between each star's GMOS instrumental magnitudes and its standardized magnitude. The average value of this offset was then applied to all stars on the GMOS frames. More work will be done in the future to further cross-check and verify this method. For stars that appeared on more than one of the master GMOS images, we averaged standardized magnitudes to get a single magnitude value.
   
   Once we obtained standardized magnitudes for the GMOS images, we constructed a CMD of the entire cluster field, and present this in the left panel of Figure \ref{all_cmd}.


\articlefiguretwo{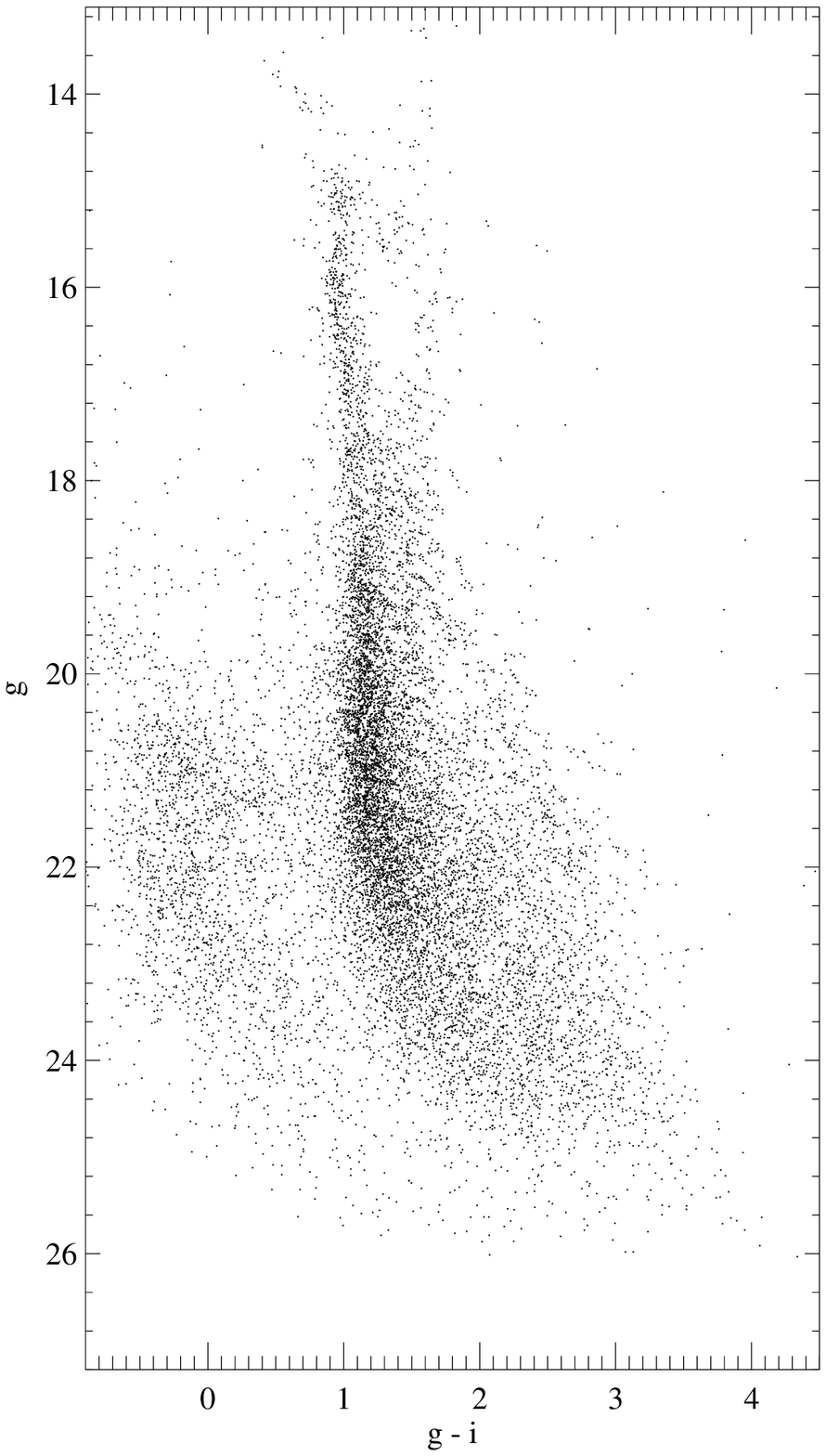}{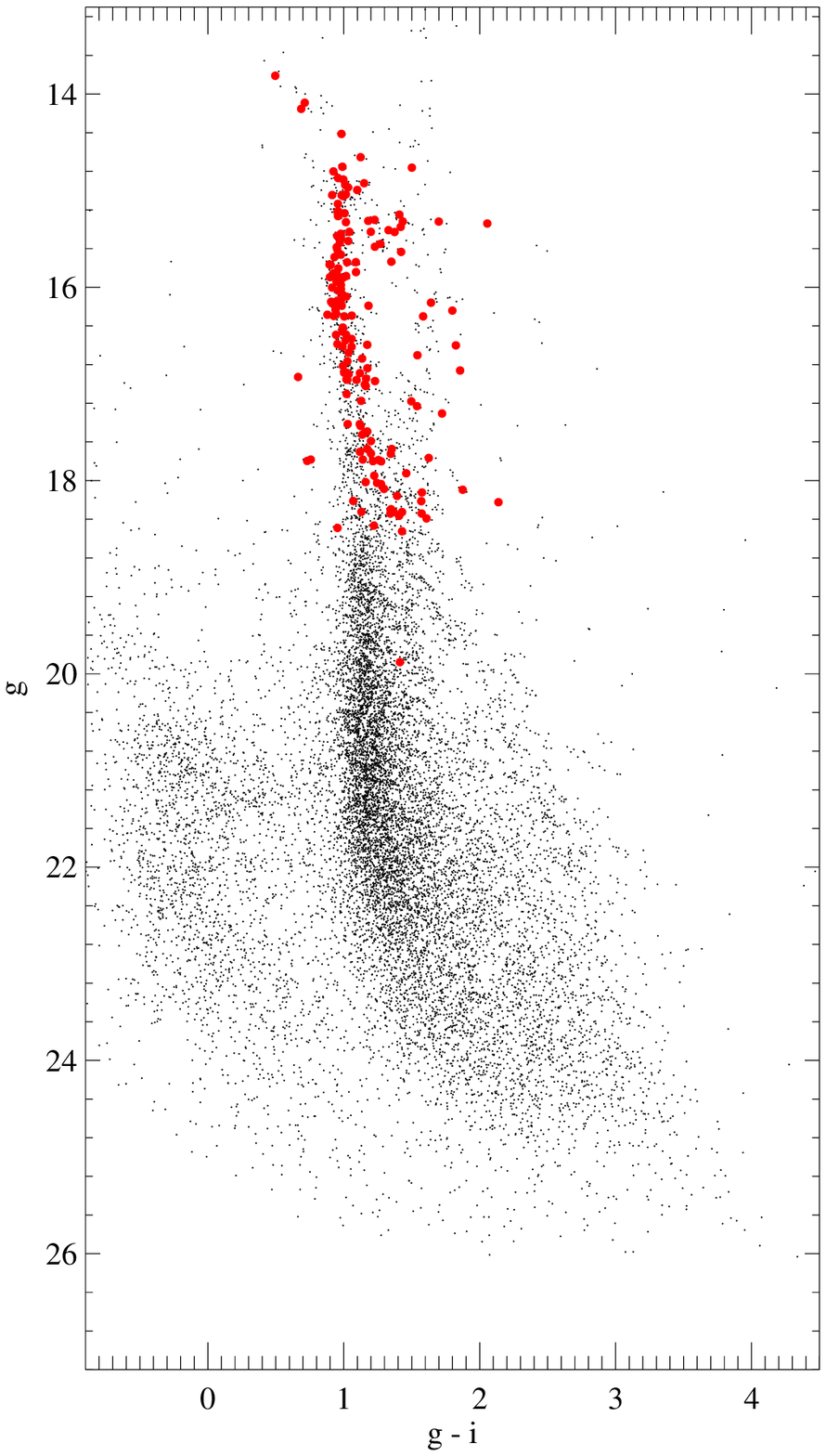}{all_cmd}{\emph{Left:} CMD of all objects in the field of NGC 6253, taken from GMOS on Gemini-South.  \emph{Right:} The same CMD with probable cluster members with $V \geq$ 18 marked. This is based on the proper motion catalog of Montalto et al. (2009).}


\section{Analysis}
\label{analysis}

Once we had a CMD of the field of NGC 6253, we first analyzed the cluster's MSTO age; then we explored the faint, blue region of the CMD looking for possible WD candidates. We describe each of these in turn here.

\subsection{Cluster MSTO Age}
\label{msto}

   Montalto et al. (2009) measured proper motions (PMs) of many bright stars ($V \geq 18$) in the field of NGC 6253. Using their PM catalog, we identified probable cluster members in our data set. We set a criteria for this of $V \geq 18$ and cluster membership probability (based on PMs, as reported by Montalto et al. 2009) $\geq$ 50\%. In the right panel of Figure \ref{all_cmd}, we again show the field of NGC 6253 with the probable cluster members marked with large dots.
   
   Once we had a clean CMD of the MSTO region of the cluster, we used the Bayesian Analysis of Stellar Evolution in Nine Variables (BASE-9; von Hippel et al. 2006; van Dyk et al. 2009; see also contributions by von Hippel et al. and Si et al. in this volume) to fit MS models of Dotter et al. (2008). We sampled on cluster age, metallicity ([Fe/H]), distance modulus, and reddening.  
   
\begin{figure}
\begin{centering}
\includegraphics[scale=0.9]{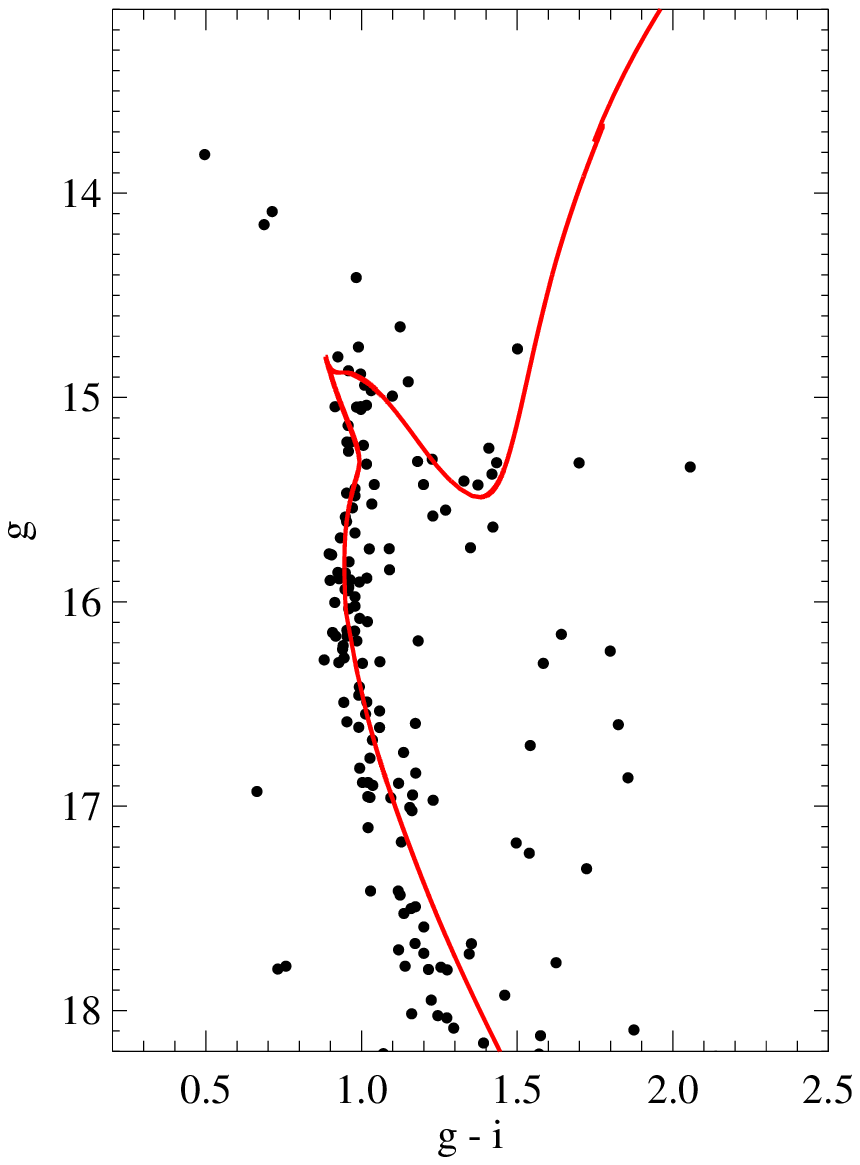}
\caption{Probable cluster members of NGC 6253. Overlaid is a Dotter et al. (2008) isochrone generated with the parameters recovered by BASE-9, including a cluster age of 4.6 Gyr.}
\label{msto_iso}
\end{centering}
\end{figure}

   By averaging posterior distributions determined by BASE-9, we found values for these cluster parameters. For this cluster, we found a MSTO age of 4.6 $\pm$ 0.2 Gyr, as well as [Fe/H] = +0.48 $\pm$ 0.06, $(m - M)_V = 11.29 \pm 0.10$, and $A_V = 0.51 \pm 0.05$ [$E(B-V) = 0.16 \pm 0.02$]. Using these values, we generated an isochrone and overlaid it on the CMD of probable cluster members. This provides a visual check of the parameters found by BASE-9. As can be seen in Figure \ref{msto_iso}, the isochrone fits the data well.



\subsection{White Dwarf Candidates}
\label{wds}

\begin{figure}
\begin{centering}
\includegraphics[scale=0.9]{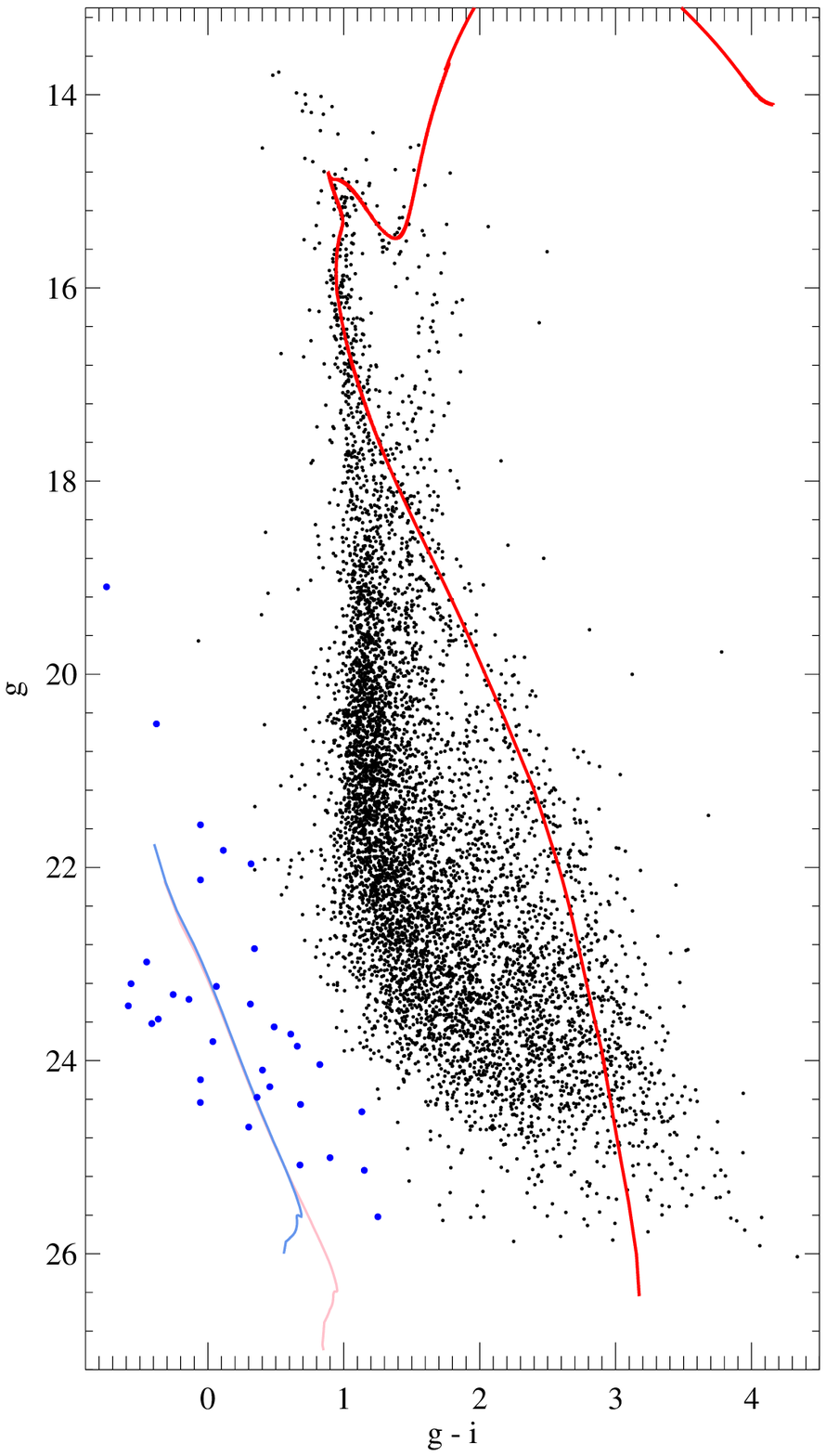}
\caption{CMD of the field of NGC 6253. Overplotted is a 4.6 Gyr MS isochrone generated with the cluster parameters discussed in Section \ref{msto}. Blue objects that were confirmed as stellar are marked with the large dots in the lower left of the CMD. Also included are two WD isochrones: one is 3.5 Gyr and one is 4.6 Gyr.}
\label{cmd_blue}
\end{centering}
\end{figure}

   To search for WD candidates on the CMD, we first made a cut on the sharpness of each object in the lower left of the CMD. This is a value determined by the DAOPHOT routines, and can be used as a criteria for star-galaxy separation. Each blue object that passed the sharpness criteria was then examined by eye, to confirm stellarity. The objects that passed both these tests are plotted as large dots on the CMD in Figure \ref{cmd_blue}. On this figure we have also overplotted a 4.6 Gyr MS isochrone, generated with the cluster parameters discussed in Section \ref{msto}. In addition, we have included two WD isochrones, generated at 3.5 Gyr (the age determined by Montalto et al. 2009) and 4.6 Gyr (the age found by Rozyczka et al. 2014, and our analysis in Section \ref{msto}). We note that the data do not go deep enough to observe the predicted WD terminus for either age.



   We have performed simulations to determine the expected number of cluster WDs in this field. Given the number of probable cluster members in a certain magnitude range on the MS (we used $g$ between 16 and 17), we simulated a cluster to match this. Assuming a cluster age of 4.6 Gyr, we would expect 167 cluster WDs in this field of view. Most of these will be faint; this simulation indicates 39 of these WDs will be brighter than the depth of our data, $g \sim 25$. We have found 35 stellar objects in the WD region of the CMD. However, we do not expect all of these objects to be cluster WDs, as this cluster suffers from a high level of field contamination due to its low Galactic latitude ($\ell \sim -6^{\circ}$). We note that we have not yet corrected for incompleteness in our data.
   
   Given the magnitude of the faintest blue objects found in this CMD, assuming cluster membership and normal C/O WD cooling models, we find a lower limit of the cluster WD age of 1.75 Gyr.
   
\section{Concluding Remarks}
\label{conclusion}

   We have presented preliminary observations and results of the metal-rich open cluster NGC 6253. Using photometry of probable cluster members observed with GMOS on Gemini-South, we have applied a Bayesian analysis and measured a cluster MSTO age of 4.6 Gyr. We identify 35 stellar objects in the WD region of the CMD, and determined a lower limit of the cluster WD cooling age of 1.75 Gyr. The bright blue stellar objects will be observed with spectroscopic follow up, with the intent of measuring the IFMR of this cluster.

\acknowledgements EJJ acknowledges support from the Department of Physics and Astronomy at Brigham Young University, and the American Astronomical Society for support to travel to the conference. FC acknowledges the support from the National Council for Scientific and Technological Development (CNPq/Brazil).


\hspace{5.2in}

\noindent
\textbf{References}\\
\hspace{5.0in}
Anthony-Twarog, B.J., et al. 2010, AJ, 139, 2034 \\
Bedin, L.R., et al. 2008a, ApJ, 678, 1279 \\
Bedin, L.R., et al. 2008b, ApJ, 679, L29 \\
Dotter, A., et al. 2008, ApJS, 178, 89 \\
Erben, T., Schirmer, M., Dietrich, J.P., et al. 2005, AN, 326, 432 \\
Garcia-Berro, E., et al. 2009, Nature, 465, 194 \\
Hansen, B.M.S. 2005, ApJ, 635, 522 \\
Montalto, M., et al. 2009, A\&A, 507, 283 \\
Rozyczka, M., et al. 2014, AcA, 64, 233 \\
Schirmer, M. 2013, ApJS, 209, 21 \\
Smith, A., et al. 2007, AJ, submitted\footnote{Data accessed via \url{http://www-star.fnal.gov/Southern_ugriz/New/index.html}} \\
Stetson, P.B. 1987, PASP, 99, 191 \\
van Dyk, D.A., et al. 2009, AnAps, 3, 117 \
von Hippel, T., et al. 2006, ApJ, 645, 1436 \\


\end{document}